# Financial Analysis of a Grid-connected Photovoltaic System in South Florida

Hadis Moradi, Amir Abtahi, and Ali Zilouchian

Florida Atlantic University, Boca Raton, FL, 33431, USA

*Abstract* —In this paper the performance and financial analysis of a grid-connected photovoltaic system installed at Florida Atlantic University (FAU) is evaluated. The power plant has the capacity of 14.8 kW and has been under operation since August 2014. This solar PV system is composed of two 7.4 kW sub-arrays, one fixed and one with single axis tracking. First, an overview of the system followed by local weather characteristics in Boca Raton, Florida is presented.
In addition, monthly averaged daily solar radiation in Boca Raton as well as system AC are calculated utilizing the PVwatts simulation calculator. Inputs such as module and inverter specifications are applied to the System Advisor Model (SAM) to design and optimize the system. Finally, the estimated local load demand as well as simulation results are extracted and analyzed.

*Index Terms*— Financial Analysis, Photovoltaic, System Advisor Model, Grid-tied Solar Unit.

## I. Introduction

While concerns over fossil fuel depletion were the primary drivers for renewable energy development in the latter part of the 20$^{th}$ Century, climate change has been the main impetus behind wind and solar use and propagation in the last 2 decades. The steep decline in the cost of large wind generators and the more dramatic drop in the cost of PV and the associated components such as inverters and power conditioners, have made these power sources competitive in numerous countries. Advances in power electronics industry have also contributed to more efficient and reliable integration of these renewable resources into electric power grids [1-3]. The electric energy industry restructuring and the introduction of the concept of a smart grid has also led to new technologies such as distributed energy resources (DER) and distributed generation (DG) become more widely utilized. Generally, DG units could be defined as a local generation units that can be directly connected to the distribution utility grid [4-8].

PV electric power generation is economically feasible and environmentally sustainable, while requiring a relatively low maintenance for its operation. PV systems are often marketed and described in terms of the DC power rating of their modules, expressed in $'s/Watt. However, the value of a grid-tied solar system is a function of the energy generated, expressed in the scale of kWh [9,10]. The PV systems can be operated in both stand-alone and grid-connected modes.
A stand–alone PV system is an autonomous system that can operate without any connection to the grid and satisfy the design load. These systems are ideal when the grid is either not available or the cost of electricity is too high. Grid-tie and hybrid systems make up the majority of PV systems currently installed and operating. Hybrid systems allow the addition of batteries for either back-up in case of utility failure, or for demand control such as peak-shaving or utilities offer Time of Use (TOU) rates.

The performance ratio (PR), often called Quality Factor (QF), is independent from the irradiation, and mostly used to compare PV system performances. The intermittent nature of solar irradiation, especially in semi-tropical regions such as Florida, does not allow any PV operator to offer a power generation guarantee. However, it's possible to predict the energy generated with a high level of certainty based on reliable solar irradiation data [11].

In Florida, a net-metering mechanism was approved in 2007 for renewable energy systems under 500 kW capacities. It allows the users to feed a portion of the electricity generated into the grid and to receive credit per kWh supplied at the same rate as the utility kWh charges. Since 2012, net metering has also been available for multi-family housing. Based on different utility companies such as FPL [20] instructions, various types of renewable energy systems such as solar energy, wind energy, biomass, ocean energy, waste heat, hydroelectric power and geothermal energy are potentially eligible for net metering.

Each tenant will pay the difference between its individual consumption and the specific PV-generated-electricity; this difference is allocated to the electric utility company (CFE) to that tenant's utility account, according to a pre-arranged share. The PV Levelized cost of energy (LCOE) has experienced a significant decrease from 2009 to 2014, which is estimated at -18.4% compound annual growth rate, even though, for the average electricity consumer PV investment is still not competitive with grid electricity prices [12].

For the PV distributed generation (PVDG), there have been numerous studies to achieve the optimum allocation of the system. As mentioned, the optimal site and size of DG reflects the maximum loss reduction and improvement in voltage profile of distribution system. Different methodologies have been developed to determine the optimum location and optimum size of the DG. These methodologies are either based on analytical tools or on optimization programming methods [13].

Photovoltaic serves as a fundamental source to harness solar energy. Accompanied with receding prices, solar leasing and other innovative financing methods PV market is spreading widely. As per statistics specified in Renewables-2013 Global Status Report by Renewable Energy Policy Network for 21st

Century, the PV industry hit a 100 GigaWatt power production in 2012 [14]. Also International Renewable Energy Agency (IRENA) estimates the global weighted average LCOE of solar PV could fall by 59 percent by 2025 from 2015 [21], which makes it more competitive compared with conventional energy solutions. For the engineering project of a large-scale PV system, the economic analysis should be performed to evaluate the profitability of the PV system to ensure the investment cost can be recovered over the life cycle. It is concluded that the main factors affecting the PV system deployment are the initial capital cost of the system, the feed-in tariff and the PV system capital cost subsidization rate [15].

The System Advisor Model (SAM) is software used in renewable energy project analysis that integrates a detailed system performance model with a financial model and cost analysis [16]. The aim of this work is to study the financial performance of a 14.8 kW solar PV system installed at FAU campus in SAM and analyze the cost parameters and obtained simulation results. The present research, evaluates the generated electricity from the 14.8kw PV system, taking in consideration the local weather conditions. The PV module temperature, wind velocity and the solar irradiations are the main parameters for PV system performance evaluation.

## II. SYSTEM OVERVIEW

The performance of a solar PV unit located at FAU in Boca Raton, Florida is studied and evaluated. This photovoltaic system consists of one fixed solar array and one tracker array. Each solar array has 12 modules in series and 2 strings in parallel. The tracking array has been designed to be installed on a North-South axis and tracking from East to West throughout the day with a broad turn range of 90°. The south facing fixed array has been designed on an East-West axis with the azimuth of 180° and tilt angle of 23° for "near optimum" annual generation [17]. The configuration of each installed array is shown in Fig .1.

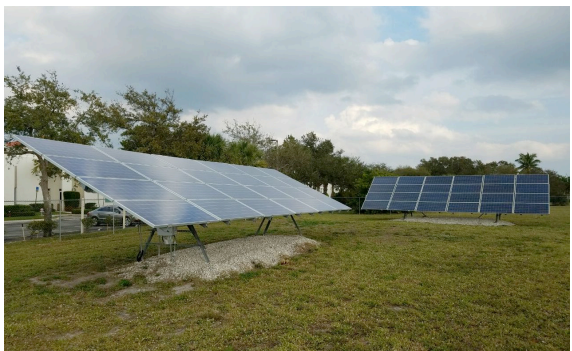

Fig. 1. The configuration of one array installed at FAU

## III. LOCAL METROLOGICAL DATA

Solar irradiation is one of the major parameters in designing solar PV systems. Also, in order to obtain the maximum power output from the PV arrays, the tilt angle must be set up correctly. Those parameters help to design the system based on the lowest solar irradiation of the year, which is going to be in the winter for south Florida location. The PVWatts software tools [22] is employed to calculate the solar irradiation based on the Typical Metrological Year (TMY) as the weather database for the location. Table. 1 shows the estimated solar radiation in (KWh/m$^2$/day), and AC Energy in (KWh) for Boca Raton, Florida, USA.

TABLE I
MONTHLY AVERAGED DAILY SOLAR RADIATION AND SYSTEM'S AC ENERGY

| Month | Solar Radiation (KWh/ /day) | AC Energy (KWh) |
|---|---|---|
| January | 4.11 | 1,463 |
| February | 4.90 | 1,589 |
| March | 5.78 | 2,010 |
| April | 4.55 | 1,556 |
| May | 5.24 | 1,820 |
| June | 4.29 | 1,440 |
| July | 5.98 | 2,033 |
| August | 4.99 | 1,736 |
| September | 4.61 | 1,535 |
| October | 3.69 | 1,304 |
| November | 5.01 | 1,700 |
| December | 4.48 | 1,590 |
| *Annual* | *4.80* | *19,776* |

By selecting the location of the project, the required data such as global irradiance, diffuse irradiance, wind speed, relative humidity, snow depth and dry bulb temperature are obtained in the model. Some of the station specifications are provided in Table 2. Also solar global irradiance (W/m$^2$) in 2014 when the system was installed at Boca Raton, FL is shown in Fig. 2.

TABLE II
LOCAL WEATHER DATA

| Weather identification | Value |
|---|---|
| Latitude | 26.37 °N |
| Longitude | -80.1 °E |
| Elevation | 7.3 m |
| Global horizontal | 5.26 kWh/m$^2$/day |
| Direct normal (beam) | 5.64 kWh/m$^2$/day |
| Diffuse horizontal | 1.68 kWh/m$^2$/day |
| Average temperature | 25.7 °C |
| Maximum snow depth | 0 cm |
| Average wind speed | 3 m/s |

Regarding the solar radiation curve in Fig .2, the solar irradiation in summertime is higher compared to the other seasons and the peak of solar potential occurs in May and minimum global irradiance happens in January.

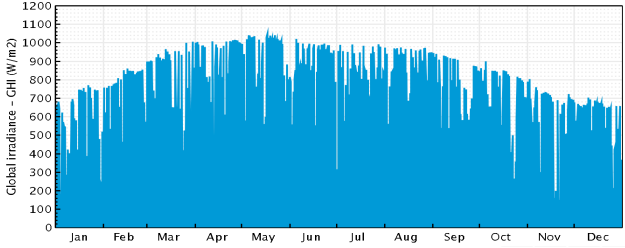

Fig. 2. Solar radiation curve in 2014, Boca Raton, FL

### IV. MODULE SPECIFICATION

The solar panel used in the system is Trina TSM 285 PA14. It composed of multicrystalline 156mm by 156mm solar cells. Each panel has 72 (6 by 12) cells and module dimensions are 1956 × 992 × 46mm. The I-V curve of the module at Standard Test Condition (STC) with total irradiance of 1000 W/m$^2$ and cell temperature of 25 °C is shown in Fig .3 (right). According to the I-V curve the maximum power point is 285.31 Wdc. Also, the electrical specifications of the module are presented in Table 3.

TABLE III
MODULE ELECTRICAL DATA

| Electrical specs | Value |
|---|---|
| Peak power watts-$P_{max}$ | 285 W |
| Power output tolerance | 0/+3 % |
| Maximum power voltage | 35.6 V |
| Maximum power current | 8.02 A |
| Open circuit voltage-$V_{oc}$ | 44.7 V |
| Short circuit current | 8.5 A |
| Module efficiency-$\eta_m$ | 14.7 % |

### V. INVERTER SPECIFICATION

In this section the employed inverter and its specifications are introduced. A Sunny Boy 7000-US inverters have been used for each solar array. The efficiency curve of the inverter is shown in Fig .3(left) and the technical data are presented in Table 4.

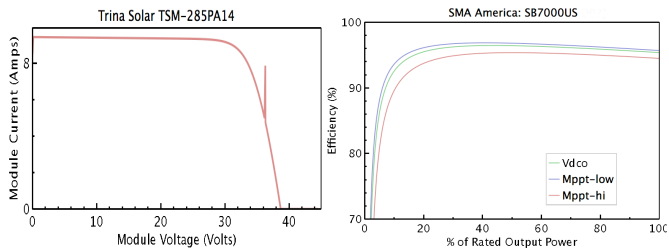

Fig. 3. (Right) Trina solar panel I-V curve at STC, (Left) Sunny Boy 7000-US efficiency curve

TABLE IV
INVERTER TECHNICAL DATA

| Technical specs | Value |
|---|---|
| Max usable PV power at STC | 8750 W |
| Max. DC power (@cos ϕ=1) | 7400 W |
| Max. DC voltage | 600 V |
| DC nominated voltage | 310 V |
| MPP voltage rate | 250-480 V |
| AC nominal power | 7000 W |
| Nominal AC voltage | 208/240/277 V |
| Max output current | 34/29/25 A |
| Efficiency | 97% |

### VI. SHADING AND LOSSES

In this section the shading, snow and possible losses, which decrease the system performance efficiency, are described. Various losses that system might experience due to the environmental conditions have significant effects on the system proper operation. Shading losses come from trees or buildings nearby or even due to self-shading from its own solar system structure and back-to-back PV rows. This system has no shading effects. Also, due to the weather condition in Boca Raton, the possibility of snow is zero, thus snow losses won't be applied to the system modeling. Other losses such as monthly soiling losses, module mismatch, diodes and connections and DC wiring are assumed 5%, 2%, 0.5% and 2% respectively. Thus the total DC power loss of 4.44% is considered as the inputs of the modeling. Also, AC wiring losses which comes from the electrical output of the inverter is assumed 1%. System performance degradation is assumed 0.5% per year, which means that total annual output of the PV system in financial modeling calculation will be degraded by half a present.

### VII. SYSTEM COSTS AND FINANCIAL PARAMETERS

In this section the direct and indirect capital costs as well as operation and maintenance cost of the system is calculated and presented in Tables 5 and 6.

TABLE V
DIRECT CAPITAL COSTS

| | Unit | kWdc/kWac per unit | $/Wdc, $/Wac | Cost |
|---|---|---|---|---|
| Module | 48 | 0.3 | 1.0 | $ 13,695.26 |
| Inverter | 2 | 7 | 0.21 | $ 2,876.01 |

| | $/Wdc | Cost |
|---|---|---|
| Balance of system equipment | 0.36 | $ 4,930.29 |
| Installer margin and overhead | 1.25 | $ 17,119.08 |
| Installation labor | 0.30 | $ 4,108.58 |

## TABLE VI
### INDIRECT CAPITAL COSTS

|  | $/Wdc | Cost |
|---|---|---|
| Permitting and environmental studies | 0.1 | $ 1,369.53 |

|  | % | Cost |
|---|---|---|
| Sale tax rate (present of direct cost) | 6.0 | $ 1,110.96 |

Based on the calculations the total direct costs, indirect costs and installed costs are $ 42,729.22, $ 2,480.49 and $ 45,209.71 respectively. Also, total installed cost per capacity is $ 3.30/Wdc. Also it can be assumed some fixed costs as operation and maintenance costs. In this study 25$/kW-yr is assumed as fixed cost by capacity and $1500 is assumed as a fixed annual cost for inverter replacements by considering manufacture warranty after 10 years.

Also in terms of financial parameters it is assumed that the installation has been done using standard loan with the inflation rate of 2% per year and Florida state tax rate of 6%. In addition, the analysis period is assumed 25 years. As incentives parameters, tax credits and direct cash incentives are considered. It is assumed that federal Investment Tax Credit (ITC) is 30% [18].

## VIII. ELECTRICITY RATES

Various electricity rate and metering structures can be employed in SAM such as TOU rates, tiered rates, demand charges and TOU tiered rates. The rates for energy charges for are available in [19]. Florida Power and Light TOU tiered energy rates for weekdays are shown in Fig 4 and Table 7. The weekends are also placed in period one. It is assumed that monthly total excess rolled over to next month bill in kWh.

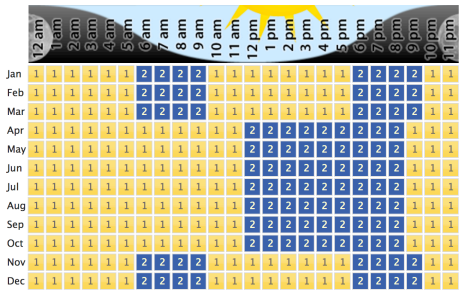

Fig. 4. FPL different periods of energy rates

## TABLE VII
### RATES FOR ENERGY CHARGES

| Period | Tier | Max. Usage | Max. Usage units | Buy ($/kWh) |
|---|---|---|---|---|
| 1 | 1 | 1000 | kWh | 0.18491 |
| 1 | 2 | 1e+38 | kWh | 0.06059 |
| 2 | 1 | 1000 | kWh | 0.17525 |
| 2 | 2 | 1e+38 | kWh | 0.06635 |

## IX. LOAD DATA

The system performance is studied under a typical load demand using load data estimator. A simple residential load model is employed in this study. We assume a building with the floor area of 4,000 sq. ft has been built in 1985. Also the cooling and heating set points is considered 64 and 72 °F. The studied hourly electric load is shown in Fig. 5. It can be seen from the load profile that the demand peak occurs in summer time when the cooling demand is high in Florida.

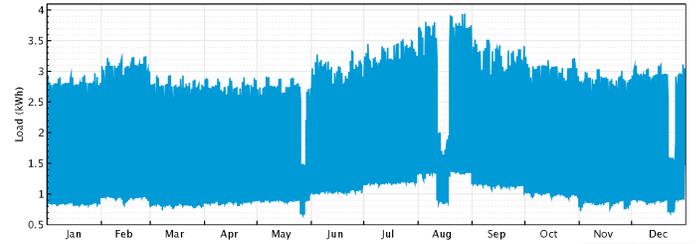

Fig. 5. Hourly electrical load demand

## X. FINANCIAL MODEL RESULTS

The model's graphical and numerical results are shown in Fig. 6 and Table 8. It is concluded from the Fig .6(a) that the system has its maximum production in springtime when the solar irradiation is high. Fig .6(c) and 6(d) show how PV production is higher in May compared to November. Also, Fig .6(b) shows the system energy generation per year, which is decreasing over the time due to the system degradation.

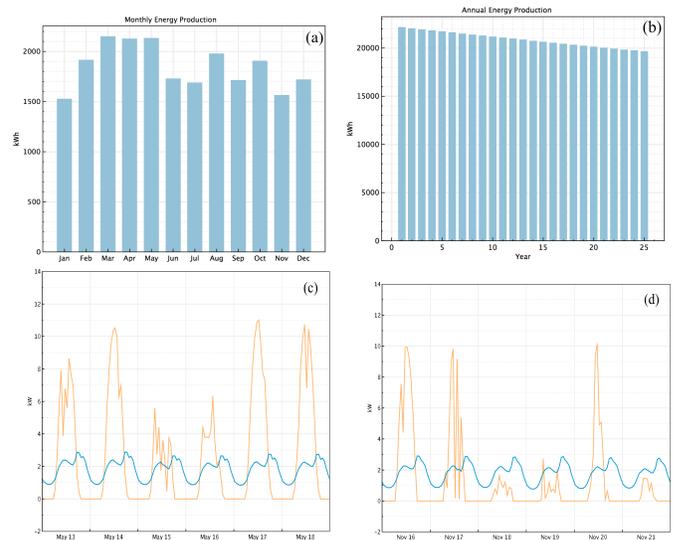

Fig. 6. (a) System monthly produced energy (b) Annual system energy generation (c) load demand vs. PV production sample in May (d) load demand vs. PV production sample in November

TABLE VIII
NUMERICAL RESULTS

| Metric | Value |
|---|---|
| Annual energy (year1) | 22,155 kWh |
| Capacity factor (year1) | 18.5% |
| Energy yield (year1) | 1,618 kWh/kW |
| Performance ratio (year1) | 0.77 |
| Eclectic bill without system (year1) | $2,637 |
| Eclectic bill with system (year1) | $112 |
| Net saving with system (year1) | $2,525 |
| Payback period | 13.9 years |
| Net Capital cost | $45,210 |

## XX. CONCLUSION

In this paper, design, implementation and financial analysis of a grid-tied PV unit were carried out to fulfill the local load demand based on corresponding meteorological parameters. The 14.8 kW system comprised of 48 modules including fixed and tracker arrays connected to two inverters to support the load at FAU campus. The simulation results showed that although installation cost is relatively high, the system would operate within disbursement period after payback time. Cost savings on electric bill would be 95.7% in the first year of operation. Additionally, the data analysis demonstrated that the unit power output is maximized in May that is around 30% more than minimum generated power in January in south Florida. To improve the system performance in terms of technical and financial applications by smoothing or shifting the profile of energy output, a set of battery storage can be integrated to existing PV system as a potential future work.